\documentclass[12pt]{article}

\usepackage{newtxtext,newtxmath}
\usepackage{graphicx}

\usepackage[letterpaper,margin=1in]{geometry}

\renewenvironment{abstract}
	{\quotation}
	{\endquotation}

\date{}

\makeatletter
\renewcommand{\fnum@figure}{\textbf{Figure \thefigure}}
\renewcommand{\fnum@table}{\textbf{Table \thetable}}
\makeatother

\usepackage{scicite}

\usepackage{url}

\def\scititle{
	Tracking molecular hydrogen formation from ionized water in real time
}
\title{\bfseries \boldmath \scititle}

\author{
	Chuan~Cheng$^{1,2,3\ast}$,
	Chi-Hong~Yuen$^{4}$,
	Eleanor~Weckwerth$^{1,2}$,
	Ian~Gabalski$^{1,5}$,\and
	Aaron~M.~Ghrist$^{1,5}$,
	Haoran~Ma$^{1,2}$,
	Andrew~J.~Howard$^{1,5}$,
	Mathew~Britton$^{1,6}$,\and
	Yunquan~Liu$^{3}$,
	Eric~Wells$^{7}$,
	Philip~H.~Bucksbaum$^{1,2,5\dagger}$
	\and
	\small$^{1}$Stanford PULSE Institute, SLAC National Accelerator Laboratory, Menlo Park, California 94025, USA.\and
	\small$^{2}$Department of Physics, Stanford University, Stanford, California 94305, USA.\and
	\small$^{3}$State Key Laboratory for Mesoscopic Physics and Frontiers Science Center for Nano-Optoelectronics,\and
    \small School of Physics, Peking University, Beijing 100871, China.\and
	\small$^{4}$Department of Physics, Kennesaw State University, Marietta, Georgia 30060, USA.\and
	\small$^{5}$Department of Applied Physics, Stanford University, Stanford, California 94305, USA.\and
	\small$^{6}$Linac Coherent Light Source, SLAC National Accelerator Laboratory, Menlo Park, California 94025, USA.\and
	\small$^{7}$Department of Physics, Augustana University, Sioux Falls, South Dakota 57197, USA.\and
	\small$^{\ast}$Corresponding author. Email: chengcc@pku.edu.cn\and
	\small$^{\dagger}$Corresponding author. Email: phbuck@stanford.edu
}

\begin{document}

% Insert the title and author list
\maketitle

% Abstract, in bold
% There are strict length limits, and not all formats have abstracts.
% Consult the journal instructions to authors for details.
% Do not cite any references in the abstract.
\begin{abstract} \bfseries \boldmath
Removing an electron from a water molecule can drive its two hydrogen atoms to pair up and depart as molecular hydrogen. However, even for this elementary reaction, the route from start to finish has remained hidden because measurements have yet to follow the electronic and nuclear motion simultaneously. Combining correlated photoelectron and ion imaging, few-femtosecond pump--probe measurements, and nonadiabatic simulations, we track the complete pathway in isolated heavy water (D$_2$O) molecules. The reaction takes an indirect route and dissociates along three distinct pathways (direct, roaming, and delayed) with formation times of about 34 and 72 femtoseconds for the direct and delayed branches. Yet bond formation requires the molecule to first break its own symmetry. Only random asymmetric motion enables the electronic-state switch at a conical intersection, joining the two hydrogen atoms before the oxygen--hydrogen bond breaks. These results establish a time-resolved picture of molecular hydrogen formation from water and provide a general strategy for linking electronic excitation to chemical outcomes in settings from radiation damage to hydrogen production.
\end{abstract}

% The first paragraph of any Science paper does NOT have a heading
% Nor is it indented
\noindent
Hydrogen transfer is among the fastest nuclear processes in molecular systems, driving chemistry from interstellar ice clouds to biological respiration\cite{marx2006proton,cuppen2024laboratory}. In the simplest aqueous unit, the isolated water molecule, this transfer manifests as proton dissociation following ionization, a process that underpins radiation damage, atmospheric chemistry, and the photocatalytic splitting of water into clean hydrogen fuel \cite{loh2020fastest,vaida2011perspective,nishioka2023photocatalytic}. Despite decades of study, the mechanism by which a water cation evolves into distinct dissociation channels remains experimentally unresolved at the molecular level, because the relevant electronic and nuclear motions are strongly coupled and occur on a few-femtosecond timescale. Tracking valence electron motion in real time has been demonstrated in atomic systems \cite{goulielmakis2010realtime,ramasesha2016real}, but in molecules the electronic state preparation and the ensuing bond-breaking nuclear rearrangement are inseparable. Resolving the complete pathway from ionization to bond cleavage therefore requires observables that are simultaneously sensitive to the initial electronic state and the transient nuclear evolution.

The water cation's three lowest electronic states ($\tilde{X}$, $\tilde{A}$, $\tilde{B}$) and their dissociation limits are well established by photoelectron spectroscopy and \textit{ab initio} theory \cite{ning2008high,truong2009threshold,sage2010velocity,cheng2020momentum,roos2018dissociations}. The $\tilde{B}$--$\tilde{A}$ conical intersection (CI) and $\tilde{A}$--$\tilde{X}$ Renner--Teller coupling mediate nonadiabatic transitions between these states during dissociation \cite{schuurman2018dynamics,suarez2018nonadiabatic,sage2010velocity,eroms2010nonadiabatic,cheng2020momentum,suarez2015nonadiabatic}, and strong-field ionization has been shown to populate multiple cationic states coherently \cite{farrell2011strong,yuen2026nonadiabatic,boguslavskiy2012multielectron}. Photoelectron--photoion coincidence measurements have correlated the ionizing electron with the kinetic energy of the resulting ion fragments, identifying which electronic states populate which dissociation channels \cite{hosaka2013correlation,zhang2016photon,cheng2020momentum}. Double ionization has been used to investigate the dissociation dynamics of water \cite{zhou2026site,gervais2009h2o2,streeter2018dissociation,reedy2018dissociation,iskandar2024tracking,pedersen2013photolysis,severt2022step} and to study H$_{2}^{+}$ or D$_{2}^{+}$ formation \cite{kechaoglou2023two,kechaoglou2020enhanced,leonard2019bond,kechaoglou2019controlling,garg2012quantum,rajgara2009strong}. However, all of these measurements are time-integrated. They fix the initial electronic state and the asymptotic fragment outcome, but the transient nuclear geometry between state preparation and bond breaking remains unobserved.

Time-resolved measurements can, in principle, access this transient regime. Coulomb explosion imaging has successfully determined absolute molecular stereochemistry and collective structural fluctuations of complex molecules \cite{vager1989coulomb,pitzer2013direct,herwig2013imaging,frasinski2016covariance,boll2022x,li2024roaming,richard2025imaging,cheng2023multiparticle,cheng2024multiparticle}, and time-resolved CEI has captured real time nuclear dynamics \cite{stapelfeldt1998time,endo2020capturing,howard2025isotope,howard2023filming}. Yet without coincident detection of the ionizing electron, these measurements alone do not reveal which electronic state launched the nuclear trajectory. Accessing the complete pathway from electronic state preparation to bond cleavage therefore requires observables that combine electron--ion correlation with femtosecond time resolution.

Here we combine photoelectron--photoion correlation spectroscopy with time-resolved ion--ion correlation imaging to track the full pathway from electron removal to D--D bond formation in D$_{2}$O. We find that D$_{2}^{+}$ production proceeds via an indirect ionization pathway initiated at the lower-lying $\tilde{A}$ state, followed by strong-field excitation to the dissociative $\tilde{B}$ state. From there, the system undergoes nonadiabatic dynamics and dissociates through three distinct dynamical modes (direct dissociation, roaming, and delayed dissociation), all governed by conical-intersection topology and symmetry breaking. Trajectory surface-hopping simulations reproduce the experimental time constants, which reveal that asymmetric nuclear motion is essential for both nonadiabatic hopping and the roaming dynamics that precede D--D bond formation, in agreement with recent observations of roaming in highly excited states in another triatomic molecule SO$_{2}$ \cite{li2024roaming}.

\subsection*{Non-Franck--Condon ionization pathway}

\begin{figure} % Do NOT use \begin{figure*}
	\centering
	\includegraphics[width=0.8\textwidth]{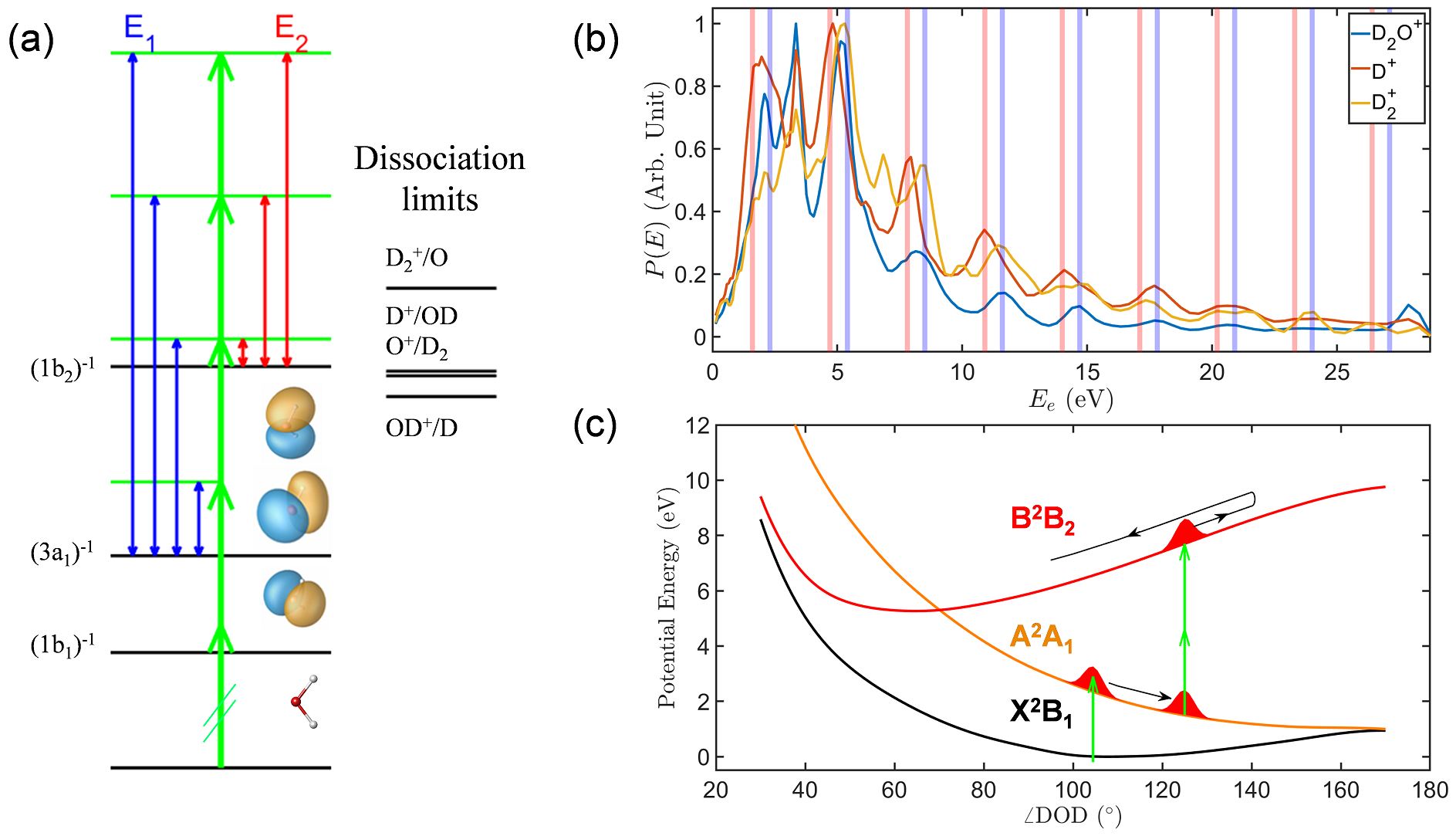}
	% Captions go below figures
	\caption{\textbf{Channel-resolved above-threshold ionization (CRATI) spectroscopy of D$_2$O.}
		(a)~Energy diagram illustrating the CRATI mechanism. Multiphoton absorption (green arrows) ionizes neutral D$_2$O from different orbitals, shifting the photoelectron energy comb according to the ionization potential of each orbital. The two resulting ATI series are labeled E$_1$ (blue arrows, ionization from HOMO--1, 3$a_1$) and E$_2$ (red arrows, ionization from HOMO--2, 1$b_2$). Cationic hole configurations [(1b$_1$)$^{-1}$, (3a$_1$)$^{-1}$, (1b$_2$)$^{-1}$] with the corresponding molecular geometries, and the dissociation limits (D$_2^+$/O, D$^+$/OD, O$^+$/D$_2$, OD$^+$/D), are indicated.
		(b)~Photoelectron kinetic energy spectra measured in coincidence with D$_2$O$^+$ (blue), D$^+$ (red), and D$_2^+$ (yellow), following ionization by a 400-nm, 40-fs laser pulse. Vertical lines mark the two ATI comb series defined in (a): E$_1$ (blue) and E$_2$ (red).
		(c)~Potential energy curves of the three lowest electronic states of D$_2$O$^+$ (X$^2$B$_1$, A$^2$A$_1$, B$^2$B$_2$) as a function of the DOD bending angle, showing the post-ionization non-Franck--Condon wave packet dynamics that drive D$_2^+$ formation via sequential excited-state population transfer (green arrows).}
	\label{fig:ele-ion correlation}
\end{figure}

The photoelectron--photoion correlation measurements reveal that D$_{2}^{+}$ formation is initiated by an indirect, non-Franck--Condon ionization pathway. Fig.~\ref{fig:ele-ion correlation}(a) shows the principle of channel-resolved above-threshold ionization (CRATI). A strong-field pulse ionizes the molecule from different orbitals, shifting the resulting photoelectron energy combs according to the ionization potential of each orbital. In the CRATI spectrum recorded with 400-nm, 40-fs pulses (Fig.~\ref{fig:ele-ion correlation}(b)), two ATI comb series are identified, corresponding to ionization from the HOMO--1 (3$a_{1}$) and HOMO--2 (1$b_{2}$) orbitals. The HOMO (1$b_{1}$) channel is suppressed at 400 nm because the photon energy (3.1 eV) makes the four-photon absorption (12.4 eV) just below the HOMO ionization potential (12.6 eV); conversely, five-photon absorption (15.5 eV) makes ionizing HOMO--1 (14.8 eV) the dominant channel \cite{zhao2014removing}. When the photoelectron spectrum is correlated with D$_{2}^{+}$, the dominant structure matches that of the parent D$_{2}$O$^{+}$ ion, produced primarily by HOMO--1 removal. The D$^{+}$ spectrum is offset by approximately 0.6 eV, consistent with HOMO--2 removal \cite{cheng2020momentum}. Weak additional peaks near 4~eV and 7~eV in the D$_{2}^{+}$-correlated spectrum are tentatively attributed to inner-valence ionization or resonant multiphoton processes (Supplementary Text). The dominant structure, however, indicates that D$_{2}^{+}$ does not emerge from direct ionization to a dissociative state; instead, the molecule must first populate the bound $\tilde{A}$ state and subsequently absorb additional photons to reach the dissociative $\tilde{B}$ state. After HOMO--1 removal, the internal energy is 14.8 eV; two additional 400-nm photons (3.1 eV each) bring the total energy to $\sim$21 eV, which exceeds the D$_{2}^{+}$/O dissociation limit of 20.5 eV. This excitation is a two-photon process driven by the strong field. At the equilibrium C$_{2v}$ geometry this excitation is less likely to happen, since the $\tilde{A}$--$\tilde{B}$ gap is larger than the single- or two-photon energy. The nuclear motion on the $\tilde{A}$ state removes this obstacle. As the molecule unbends toward the linear Renner--Teller geometry, the $\tilde{A}$--$\tilde{B}$ gap widens into resonance with two 400-nm photons (Fig.~\ref{fig:ele-ion correlation}(c)). This sequential ionization--excitation mechanism, in which nuclear motion precedes the second excitation step, is the defining feature of the non-Franck--Condon pathway\cite{zhao2014removing,li2019multiorbital}.

Further evidence for the non-Franck--Condon feature emerges from the time-resolved pump--probe data. The ion--ion correlation measurements employed a pair of 6-fs, 800-nm pulses with orthogonal linear polarizations. The orthogonal polarization geometry avoids optical interference between the two pulses; with parallel polarizations, the resulting oscillations of the strong-field ionization yield would overwhelm the weaker delay-dependent dissociation dynamics. The pump--probe delay was scanned from $-20$~fs to $+100$~fs in 2-fs steps. At each time step, ion--ion covariance technique was used to extract the (D$_{2}^{+}$, O$^{+}$) coincidence channel from the multi-hit data, rejecting background events in which the two ions originate from different molecules (see Supplementary Text for details of the covariance and cumulant mapping analysis). In the KER--delay map of the (D$_{2}^{+}$, O$^{+}$) channel (Fig.~\ref{fig:ion-ion correlation}(a)), the signal is suppressed at $t = 0$ relative to non-zero delays. For a direct, Franck--Condon dissociation, the strongest signal should appear at zero delay, where the pump pulse has maximum intensity. The observed suppression indicates that the D$_{2}^{+}$ pathway requires nuclear motion away from the Franck--Condon region before the dissociation coordinate can be accessed. The same suppression is observed in the (HD$^{+}$, O$^{+}$) channel (Fig.~\ref{fig:ion-ion correlation}(b)), but not in the (H$_{2}^{+}$, O$^{+}$) channel (Fig.~\ref{fig:ion-ion correlation}(c)), where the lighter isotope's faster dynamics allow partial Franck--Condon dissociation even within the 6-fs pulse envelope. This isotopic difference confirms that the suppression is a dynamical effect rather than an artifact. Furthermore, this indirect pathway also explains the high intensity threshold ($\sim$10$^{15}$W/cm$^{2}$) reported previously \cite{rajgara2009strong,kechaoglou2020enhanced}. Because the excitation and dissociation steps are separated in time, the nuclear rearrangement proceeds only after the multiphoton excitation is complete, and both steps must be completed within the same pulse, which is achieved only at sufficiently high field strengths \cite{howard2023filming,cheng2023multiparticle}.

\subsection*{Three distinct dissociation modes}

\begin{figure} % Do NOT use \begin{figure*}
	\centering
	\includegraphics[width=0.8\textwidth]{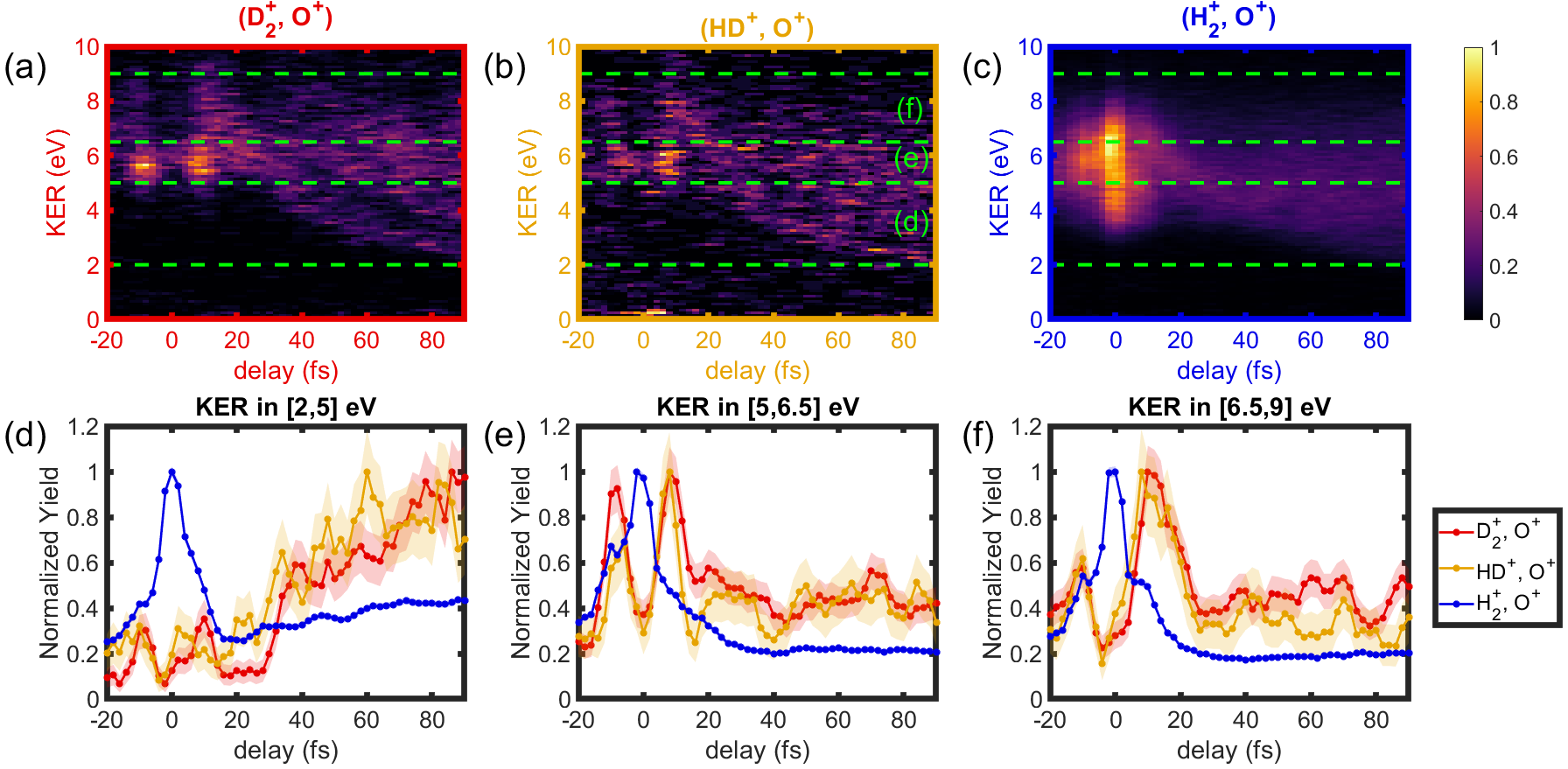}
	% Captions go below figures
	\caption{\textbf{Kinetic energy release (KER)--resolved ion--ion coincidence dynamics following strong-field ionization of water isotopologues.}
		(a-c)~Two-dimensional maps of ion--ion coincidence yield as a function of pump--probe delay and KER for (D$_2^+$, O$^+$) (a), (HD$^+$, O$^+$) (b), and (H$_2^+$, O$^+$) (c), following ionization by two 800-nm, 6-fs pulses. Green dashed lines mark the KER integration windows for the lineouts below.
		(d-f)~Delay-dependent normalized yield integrated over KER ranges [2, 5]~eV (d), [5, 6.5]~eV (e), and [6.5, 9]~eV (f). Notably, the (D$_2^+$, O$^+$) and (HD$^+$, O$^+$) channels exhibit suppressed yield at zero delay, inconsistent with the field-driven ionization maximum expected at temporal overlap. This suppression indicates that the dissociation proceeds from a non--Franck--Condon initial wave packet, in agreement with the photoelectron spectrum in Figure~\ref{fig:ele-ion correlation}.}
	\label{fig:ion-ion correlation}
\end{figure}

The pump--probe data show several distinct features and modes for D$_{2}^{+}$ formation, each with a characteristic timescale and KER signature. Fig.~\ref{fig:ion-ion correlation}(a) shows the KER--delay map for the (D$_{2}^{+}$, O$^{+}$) channel, where three features stand out, namely the suppression of signal at $t = 0$ in the 2--5 eV range, the rise of population in this band with increasing delay (corresponding to the increasing O--D$_{2}$ distance), and an enhanced signal at $\pm 8$ fs and KER~$\approx$~4.5 eV. The corresponding 1D lineouts (Fig.~\ref{fig:ion-ion correlation}(d)-(f)) show the time-dependent population in selected KER bands. In the 5--6.5 eV range (Fig.~\ref{fig:ion-ion correlation}(e)), the population exhibits a dip at $t = 0$, then rises and oscillates with a period of $\sim$20--30 fs. In the 6.5--9 eV range (Fig.~\ref{fig:ion-ion correlation}(f)), the population shows a complementary oscillation, indicating that the dissociating wave packet is moving coherently between two regions of the potential-energy surface. The complementary oscillation, in which a dip in one band coincides with a peak in the other, is consistent with coherent population transfer between the adiabatic surfaces mediated by the conical intersection, the high- and low-KER bands reporting the population on the $\tilde{B}$ and $\tilde{A}$ surfaces, respectively. The oscillation period reflects the round-trip time of the wave packet between the CI seam and the upper-state well. In the 2--5 eV range (Fig.~\ref{fig:ion-ion correlation}(d)), the population rises monotonically with delay, reflecting the increasing O--D$_{2}$ distance as the molecule dissociates. The enhanced signal at $\pm$8 fs and KER~$\approx$~4.5 eV does not correspond to any of the simulated trajectories; rather, it arises from a breakdown of the point-charge approximation used to convert KER into internuclear distance when the O--D$_{2}$ separation is small (see Supplementary Text). At these early delays, the transient molecular geometry is still compact, and the KER--$R$ mapping is not valid, producing an artificial enhancement that vanishes once the molecule has moved beyond the critical distance. The remaining features are captured by trajectory surface-hopping (TSH) simulations of the full ensemble (Supplementary Text), which reveal their microscopic origin and resolve the dissociation into three distinct modes (direct dissociation, roaming, and delayed dissociation), as detailed below.

\begin{figure} % Do NOT use \begin{figure*}
	\centering
	\includegraphics[width=0.6\textwidth]{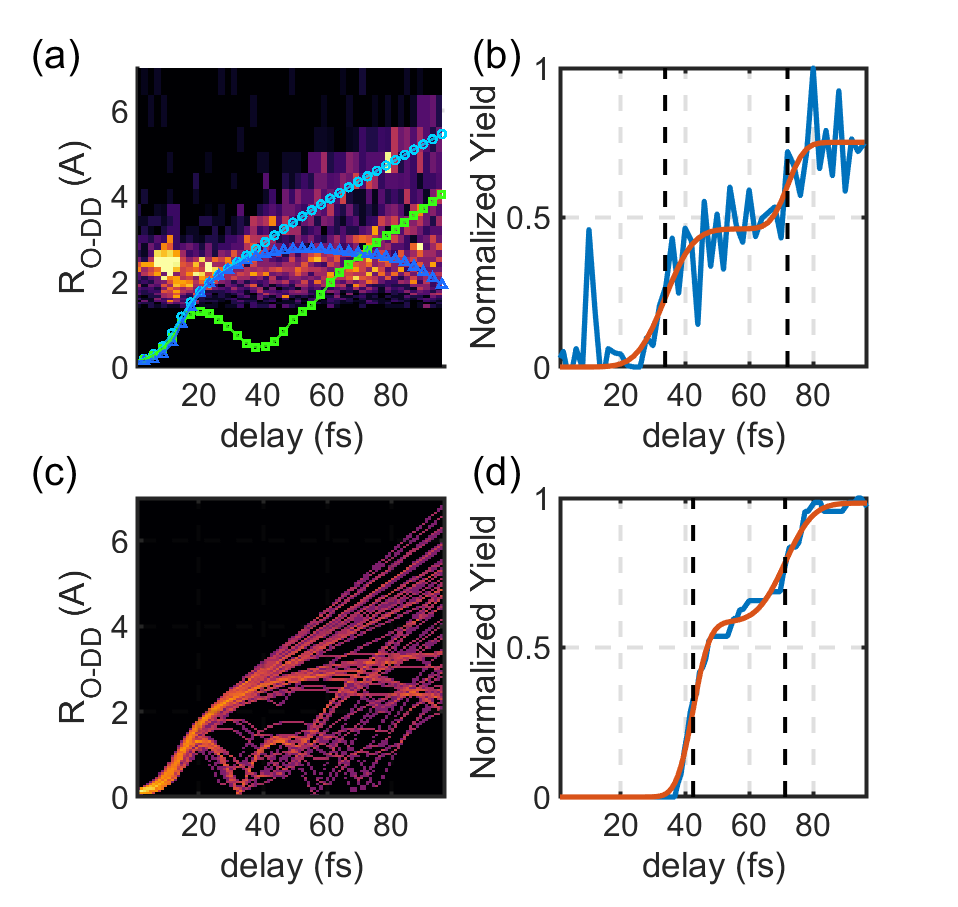}
	% Captions go below figures
	\caption{\textbf{Comparison of simulated and experimental D$_2^+$ dissociation dynamics.}
		(a)~Experimentally inferred O--DD distance $R_{\mathrm{O\text{--}DD}}$--delay map obtained via the CEI mapping $R_{\mathrm{O\text{--}DD}} = 1/\mathrm{KER}$, with the three simulated trajectories from (c) overlaid for comparison (direct dissociation, cyan circles; roaming, blue triangles; delayed dissociation, green squares).
		(b)~Normalized experimental yield integrated above $R_{\mathrm{O\text{--}DD}} = 2.6$~\AA, fitted with two complementary error functions (Eq.~S1) centered at $\mu_1 = 33.8$~fs ($\sigma_1 = 9.8$~fs) and $\mu_2 = 72$~fs ($\sigma_2 = 5.9$~fs).
		(c)~Ab initio trajectory surface hopping simulation of D$_2^+$ formation, showing $R_{\mathrm{O\text{--}DD}}$ as a function of pump--probe delay. The density map represents the ensemble of 69 trajectories that ends up with D$_2^+$ formation.
		(d)~Normalized yield obtained by integrating the simulated trajectories above $R_{\mathrm{O\text{--}DD}} = 2.6$~\AA, fitted with two complementary error functions centered at $\mu_1 = 42.2$~fs ($\sigma_1 = 4.6$~fs) and $\mu_2 = 71.5$~fs ($\sigma_2 = 15.2$~fs).
		}
	\label{fig:fitD2_theory}
\end{figure}

\begin{figure} % Do NOT use \begin{figure*}
	\centering
	\includegraphics[width=0.8\textwidth]{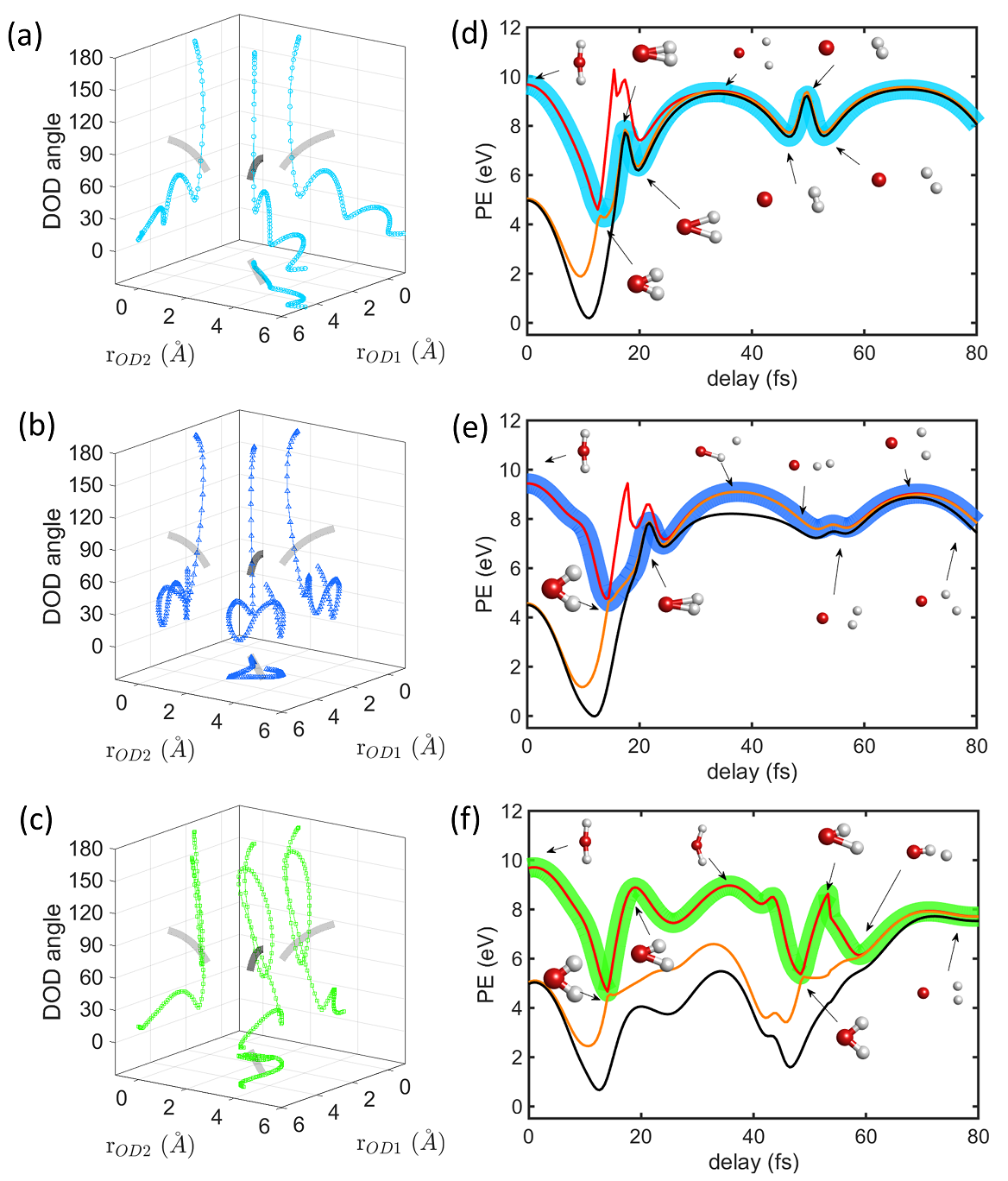}
	% Captions go below figures
	\caption{\textbf{Snapshots of three characteristic trajectories.}
		(a-c)~Evolution of the three representative trajectories from the $\tilde{B}$ state along the DOD angle and the two O--D internuclear distances (OD1, OD2). The grey shaded region marks the $\tilde{B}$--$\tilde{A}$ conical intersection seam.
		(d-f)~Transient energy of the three trajectories (shaded regions) superimposed on the first three cationic potential-energy surfaces (X$^2$B$_1$ (black), A$^2$A$_1$ (orange), and B$^2$B$_2$ (red)) computed at each instantaneous geometry.}
	\label{fig:simulation_more}
\end{figure}

The first mode is a direct dissociation following nonadiabatic hopping at the $\tilde{B}$--$\tilde{A}$ conical intersection. The wave packet, launched on the $\tilde{B}$ state, bends inward and encounters the CI seam at approximately 12 fs. A nonadiabatic hop transfers population to the lower $\tilde{A}$ surface, leading to monotonic dissociation. The experimentally inferred O--D$_{2}$ distance--delay map (Fig.~\ref{fig:fitD2_theory}(a)) shows the rising population of the direct-dissociation channel, and the error-function fit (Supplementary Text) to the integrated yield gives a first rise centered at 33.8~fs with a width of 9.8~fs (Fig.~\ref{fig:fitD2_theory}(b)). The ab initio trajectory surface-hopping simulation produces a similar ensemble density map (Fig.~\ref{fig:fitD2_theory}(c)), with a corresponding fit yielding a first rise centered at 42.2~fs and a width of 4.6~fs (Fig.~\ref{fig:fitD2_theory}(d)). The difference in rise time likely reflects the finite probe duration and the simplified initial conditions in the simulation, which samples only thermal velocities. The direct-dissociation trajectory is shown in Fig.~\ref{fig:simulation_more}(a), with its energy evolution in Fig.~\ref{fig:simulation_more}(d).

The second mode involves roaming dynamics in which the molecule hops at the CI but then turns back toward the oxygen, executing large-amplitude D--D--O motion before eventual dissociation. The roaming trajectory in Fig.~\ref{fig:simulation_more}(b) shows characteristic turning-point behavior. After the initial hop at $\sim$12 fs, the O--D$_{2}$ distance increases to $\sim$3.0 \AA, then decreases back to $\sim$1.5 \AA\ as one deuterium temporarily orbits the other, and only increases again after the second turning point. The corresponding energy evolution is shown in Fig.~\ref{fig:simulation_more}(e). This large-amplitude motion is the signature of a frustrated dissociation, in which the molecule has gained enough kinetic energy to leave the conical intersection region but not enough to overcome the effective centrifugal barrier associated with the O--D$_{2}$ bond at small bend angles. The orbital snapshots in Fig.~\ref{fig:electronDensity} show that during the roaming phase, the hole density reorganizes to build a transient bonding interaction between the two deuteriums while a residual bond to the oxygen is still maintained. This dual bonding character, in which the new D--D bond is established before the O--D$_2$ bond is fully broken, is the electronic hallmark of roaming-mediated molecular formation, distinguishing it from a simple direct dissociation. The same motion can be identified in the experimental data. The simulated roaming trajectory overlaid on the measured distance--delay map (Fig.~\ref{fig:fitD2_theory}(a), blue triangles) passes through the enhanced population at intermediate O--D$_2$ separations ($\sim$2--3~\AA), assigning this lingering band between the first and second rises to the roaming mode.

The third mode is a delayed dissociation in which the trajectory remains on the upper $\tilde{B}$ surface after the first CI encounter, bounces off a high barrier, and dissociates only after a second passage through the CI region. The trajectory in Fig.~\ref{fig:simulation_more}(c) shows a bend--unbend--bend cycle. The DOD angle decreases to $\sim 90^{\circ}$ at the first CI encounter, then increases back to $\sim 170^{\circ}$ as the trajectory reflects from the barrier, and finally decreases again to $\sim 90^{\circ}$ at the second encounter, where the trajectory adiabatically exits to the D$_{2}^{+}$ + O dissociation limit. The corresponding energy evolution is shown in Fig.~\ref{fig:simulation_more}(f). This pathway produces the longest time constant, with the error-function fit giving a second rise centered at 72~fs with a width of 5.9~fs. The coexistence of these three modes, direct, roaming, and delayed, is the hallmark of a conical-intersection-driven reaction, where the branching between pathways is controlled by the geometry at which the hop occurs and the subsequent momentum along the seam direction.

\subsection*{Symmetry breaking as the prerequisite for bond formation}

\begin{figure} % Do NOT use \begin{figure*}
	\centering
	\includegraphics[width=0.8\textwidth]{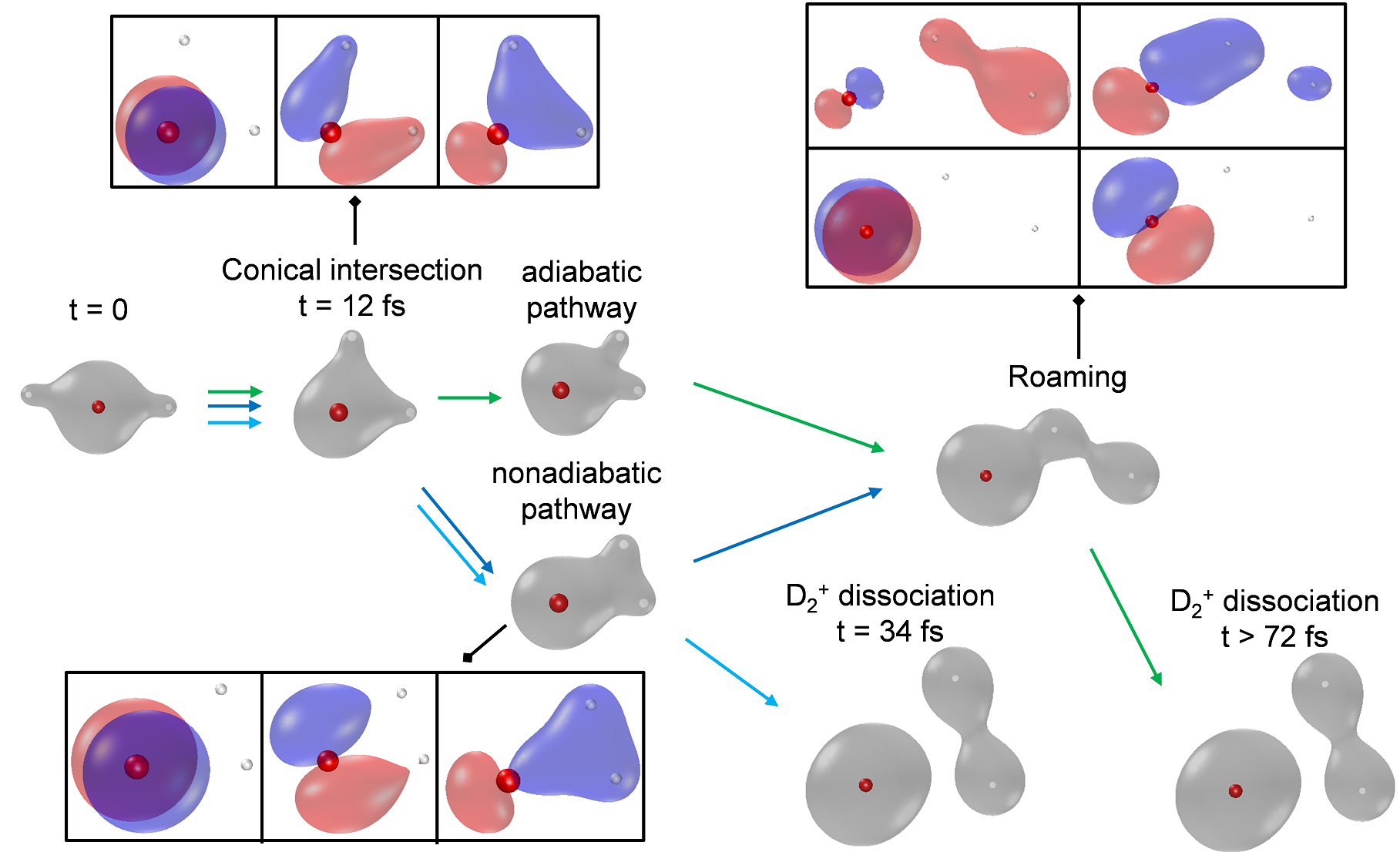}
	% Captions go below figures
	\caption{\textbf{Electron-density evolution along the branching pathways to D$_2^+$ formation.}
		Center panels show the total electron density (grey isosurfaces, with the molecular geometry embedded) at $t = 0$, at the $\tilde{B}$--$\tilde{A}$ conical intersection ($t \approx 12$~fs), and along the two subsequent pathways. The adiabatic branch (green arrows) proceeds through roaming geometries and does not reach dissociation within the propagation window ($t > 72$~fs), whereas the nonadiabatic branch (blue arrows) yields direct D$_2^+$ dissociation at $t \approx 34$~fs. Boxed insets magnify the active (singly occupied) molecular orbital at key geometries, with red and blue isosurfaces denoting its opposite wavefunction phases. As the asymmetric distortion lowers the molecular symmetry and mixes the 1$b_{2}$ and 3$a_{1}$ orbitals, the active orbital evolves continuously from O--D bonding character into a D--D bonding shape, establishing the $\sigma$ bond between the two deuteriums before the O--D$_2$ bond is fully broken. In both branches, the asymmetric distortion of the electron density, which breaks the equivalence of the two O--D bonds, is essential for reaching the dissociation asymptote.}
	\label{fig:electronDensity}
\end{figure}

The trajectory analysis reveals that asymmetric nuclear motion is essential for both nonadiabatic hopping and roaming dynamics. Under C$_{2v}$ symmetry, the nonadiabatic coupling between the $\tilde{A}$ ($^{2}A_{1}$) and $\tilde{B}$ ($^{2}B_{2}$) states vanishes by symmetry. The coupling matrix element $\langle\tilde{A}|\partial/\partial Q|\tilde{B}\rangle$ is nonzero only if the triple direct product $A_{1} \otimes \Gamma(Q) \otimes B_{2}$ contains the totally symmetric representation $A_{1}$, which requires the coupling coordinate $Q$ to transform as $b_{2}$. In the water cation, the only $b_{2}$ nuclear degree of freedom is the asymmetric stretch, so the two states remain uncoupled at any symmetric geometry. The bending mode preserves C$_{2v}$ symmetry and therefore plays a different role. It tunes the energy gap between the $\tilde{A}$ and $\tilde{B}$ states, bringing them to near degeneracy as the DOD angle decreases toward the conical-intersection seam. Hopping thus requires a symmetry-breaking distortion along the asymmetric stretch at precisely the moment when the bending motion brings the two states together. The symmetry breaking is stochastic. Even though all trajectories start from the same symmetric geometry, thermal velocity fluctuations at 300 K cause some trajectories to pick up asymmetric motion within the first $\sim$12 fs, precisely when they encounter the CI. The orbital snapshots (Fig.~\ref{fig:electronDensity}) show how the electronic density reorganizes during this process. At the equilibrium geometry, the active orbital on the $\tilde{B}$ state is the 1$b_{2}$ orbital, which has bonding character along the O--D bonds but no density between the two deuteriums. As the molecule bends and the asymmetric stretch develops, this orbital mixes increasingly with the 3$a_{1}$ orbital, which carries the hole on the $\tilde{A}$ state; the same $b_{2}$ asymmetric distortion that couples the two electronic states enables this orbital mixing. The degree of mixing is governed by the magnitude of the asymmetric displacement. At small distortions the two states approach the conical intersection, while at larger distortions the adiabatic surfaces separate and the transition probability drops. Consequently, hopping is most probable at intermediate distortions where seam geometry and momentum along the seam direction are optimally matched.
Near the conical intersection, the energy ordering of the (1$b_{2}$)$^{-1}$ and (3$a_{1}$)$^{-1}$ hole configurations is inverted, and the asymmetric distortion lowers the molecular symmetry to C$_{s}$, in which the 1$b_{2}$ and 3$a_{1}$ orbitals transform identically ($a^\prime$) and mix. The singly occupied orbital therefore evolves continuously from O--D bonding character into a D--D bonding shape, so that a $\sigma$ bond between the two deuteriums is established electronically before the O--D$_{2}$ bond is fully broken. This evolution is directly visible in Fig.~\ref{fig:electronDensity}, where the magnified insets track the active molecular orbital from the symmetric equilibrium geometry, through the conical intersection at $\sim$12~fs, and into the two branching pathways.

These microscopic features manifest in three distinct dissociation modes as shown in Fig.~\ref{fig:simulation_more} and Fig.~\ref{fig:electronDensity}. Trajectories that develop the optimal intermediate distortion undergo efficient hopping at the first CI encounter and proceed directly to dissociation, giving the fastest mode. Trajectories that experience even stronger symmetry breaking, in which one deuterium remains near the oxygen ($r_{\text{OD}} \approx 1.2$ \AA) while the other moves to $r_{\text{OD}} \approx 2.4$ \AA, create a transient O--D--D geometry in which the 3$a_{1}$ and 1$b_{1}$ orbitals become nearly degenerate (the $\tilde{A}$--$\tilde{X}$ Renner--Teller pair). This degeneracy traps one deuterium near the oxygen while the other orbits around it, creating the frustrated dissociation that characterizes the roaming dynamics. Trajectories whose initial distortion is too small to hop at the first encounter remain on the upper $\tilde{B}$ surface, reflect from the barrier, and dissociate only after a second CI passage, producing the delayed mode.

The symmetry argument also explains the low D$_{2}^{+}$ yield. Only trajectories that develop the right amount of asymmetric motion, enough to enable hopping and roaming but not so much as to trigger fragmentation into D$^{+}$ or OD$^{+}$ can form D$_{2}^{+}$. This narrow window explains why only 69 out of 500 simulated trajectories satisfy the D$_{2}^{+}$ criteria and why the experimental branching ratio is similarly small. The low yield is not a statistical artifact but a physical signature of the symmetry requirements for bond formation.

\subsection*{Conclusions}
These results establish a complete, time-resolved picture of hydrogen formation from ionization to new bond formation. By correlating the ionizing electron with the ion fragments, we have shown that the reaction proceeds via an indirect pathway initiated at the lower-lying $\tilde{A}$ state, followed by two-photon excitation to the dissociative $\tilde{B}$ state, and nonadiabatic population transfer through three distinct dynamical modes. The central role of symmetry breaking in enabling both the nonadiabatic hopping and the roaming dynamics demonstrates that asymmetric nuclear motion is a prerequisite for D--D bond formation in the water cation. This mechanism explains why previous strong-field measurements reported high intensity thresholds and low branching ratios. The reaction requires a specific sequence of ionization, bending, and additional photon absorption, all of which are sensitive to the laser field strength and duration. The isotope-dependent dynamics and the three-mode branching are consistent with a conical-intersection topology in which the seam geometry and the momentum distribution at the crossing point control the partitioning between direct, roaming, and delayed pathways. The correlation of electronic and nuclear observables demonstrated here provides a general strategy for tracking nonadiabatic dynamics in molecular systems. Beyond spectroscopy, this capability points toward quantum chemical control. Because the branching between pathways is governed by the nuclear geometry at the conical intersection, the reaction outcome can in principle be steered by manipulating the nuclear wave packet before it reaches the seam. Chirped or two-color pulse sequences could, for example, steer the arrival time and momentum of the wave packet at the CI, selectively enhancing or suppressing individual dissociation channels, and the isotope dependence suggests mass-selective control of the branching ratios. These possibilities open a route toward laser-controlled bond rearrangement in the simplest aqueous unit.

%%%%%%%%%%%%%%%% MAIN TEXT FIGURES %%%%%%%%%%%%%%%

%%%%%%%%%%%%%%%% REFERENCES %%%%%%%%%%%%%%%

\clearpage % Clear all remaining figures and tables then start a new page

% The list of references goes after the main text and before the acknowledgements
\bibliography{refs.bib}
\bibliographystyle{sciencemag}

%%%%%%%%%%%%%%%% ACKNOWLEDGEMENTS %%%%%%%%%%%%%%%

\section*{Acknowledgments}
We thank colleagues at the Stanford PULSE Institute and SLAC National Accelerator Laboratory for their support and assistance.
\paragraph*{Funding:}
C.C., E. Weckwerth, I.G., A.M.G., H.M., A.J.H., and P.H.B.
were supported by the National Science Foundation. 
A.J.H. was additionally supported under a Stanford
Graduate Fellowship as the 2019 Albion Walter Hewlett
Fellow. A.M.G. was additionally supported by an NSF
Graduate Research Fellowship. M.B. was supported by the Linac Coherent Light Source, SLAC National Accelerator Laboratory, which is supported by the US Department of Energy, Office of Science, Office of Basic Energy Sciences, under contract no. DE-AC02-76SF00515. E. Wells was supported by National Science Foundation grant number PHY-2607920.
\paragraph*{Author contributions:}
C.C., E. Wells and P.H.B. conceived the project. C.C., E. Weckwerth, I.G., A.M.G. and E. Wells. performed the experiments. C.C. analyzed the data. C.C. performed the theoretical calculations. C.C., E. Weckwerth, I.G., A.M.G., A.J.H. and M.B. contributed to the experimental apparatus. A.J.H. and M.B. performed the initial experimental demonstration. C-H.Y. and Y.L. contributed to the theoretical calculations. C.C., E. Weckwerth, I.G., A.M.G., H.M., M.B., Y.L., E. Wells and P.H.B. discussed the results. C.C. wrote the manuscript with input from all co-authors.
\paragraph*{Competing interests:}
There are no competing interests to declare.
\paragraph*{Data and materials availability:}
All data needed to evaluate the conclusions in the paper are present in the paper and/or the Supplementary Text. Additional data and analysis code are available from the corresponding authors upon request.

%%%%%%%%%%%%%%%% SUPPLEMENT LIST %%%%%%%%%%%%%%%

% List the contents of your Supplementary Materials, including the numbers of any
% supplementary figures, tables, external data files etc. and any references that are
% cited only in the supplement. In this example, refs. 7-8 are cited only in the supplement.
% Fill out your numbers accordingly and delete any lines that aren't applicable.
\subsection*{Supplementary materials}
Materials and Methods\\
Supplementary Text\\
Figures S1 to S5\\
Table S1

%%%%%%%%%%%%%%%% END OF MAIN TEXT %%%%%%%%%%%%%%%

\newpage

%%%%%%%%%%%%%%%% START OF SUPPLEMENT %%%%%%%%%%%%%%%

% Figures, tables, equations and pages in the supplement are numbered S1, S2 etc.
\renewcommand{\thefigure}{S\arabic{figure}}
\renewcommand{\thetable}{S\arabic{table}}
\renewcommand{\theequation}{S\arabic{equation}}
\renewcommand{\thepage}{S\arabic{page}}
\setcounter{figure}{0}
\setcounter{table}{0}
\setcounter{equation}{0}
\setcounter{page}{1} % not 0 as \newpage already started a supplementary page
% References continue the numbering from the main text.

%%%%%%%%%%%%%%%% SUPPLEMENT TITLE PAGE %%%%%%%%%%%%%%%

\begin{center}
\section*{Supplementary Materials for\\ \scititle}

% Author list for the supplement
% Indicate the corresponding authors, but do NOT include institutions here
% It would be nice if the template auto-generated this, but doing so is complicated...
Chuan~Cheng$^{\ast}$,
Chi-Hong~Yuen,
Eleanor~Weckwerth,
Ian~Gabalski,
Aaron~M.~Ghrist,
Haoran~Ma,
Andrew~J.~Howard,
Mathew~Britton,
Yunquan~Liu,
Eric~Wells,
Philip~H.~Bucksbaum$^{\dagger}$\\ % we're not in a \author{} environment this time, so use \\ for a new line
\small$^{\ast}$Corresponding author. Email: chengcc@pku.edu.cn\\
\small$^{\dagger}$Corresponding author. Email: phbuck@stanford.edu
\end{center}

% Fill out the numbers for each type of supplementary material,
% and delete any lines that aren't applicable.
% These are just example numbers that don't match the rest of this template.
\subsubsection*{This PDF file includes:}
Materials and Methods\\
Supplementary Text\\
Figures S1 to S5\\
Table S1

\newpage

%%%%%%%%%%%%%%%% MATERIALS AND METHODS %%%%%%%%%%%%%%%

\subsection*{Materials and Methods}

The photoelectron--photoion correlation measurements were performed with 400-nm, 40-fs pulses at 1~kHz. The beam was generated by second-harmonic generation (SHG) of a 40-fs, 800-nm Ti:sapphire amplifier (1~mJ per pulse), delivering approximately 20~mW of 400-nm light. The beam was focused into the interaction region by a 10-cm-focal-length broadband dielectric mirror. The vacuum chamber maintained a base pressure of $3\times10^{-9}$~Torr. Deuterated water (D$_{2}$O) was introduced through a precision leak valve, with the operating pressure capped at $7\times10^{-9}$~Torr to ensure moderate count-rate conditions for covariance analysis. The peak intensity of the 400-nm pulse was estimated to be approximately $4\times10^{14}$~W~cm$^{-2}$ from the pulse energy and a calculated focal spot size, assuming a Gaussian spatial profile and a sech$^{2}$ temporal envelope ($I_{0} = E_{\text{pulse}}/\pi w_{0}^{2}\tau$).

The time-resolved ion--ion correlation measurements employed a pair of 6-fs, 800-nm pulses with orthogonal linear polarizations. The 800-nm beam was first spectrally broadened in a 2.5-m differentially pumped hollow-core fiber (HCF) filled with argon at the pressure of 26~psi, then compressed by broadband chirped mirrors to a full-width-at-half-maximum (FWHM) duration of 6~fs, characterized by a d-scan device. A Michelson interferometer with waveplate--polarizer combinations generated two pulses of equal intensity with perpendicular polarizations, with the pump pulse polarized along the $x$-axis (parallel to the detector plane) and the probe along the $z$-axis (time-of-flight axis). The cross-polarized configuration suppresses optical interference between the two arms. The pump--probe delay was scanned from $-20$~fs to $+100$~fs in 2-fs steps. The 800-nm beam was focused to a spot size (1/$e^{2}$ radius) of $w_{0} \approx 8$~$\mu$m, giving a peak intensity of approximately $1\times10^{15}$~W~cm$^{-2}$ per pulse ($I_{0} = E_{\text{pulse}}/\pi w_{0}^{2}\tau$), again with an estimated $\pm$30\% uncertainty.

Both experiments shared a home-built velocity-map-imaging (VMI) spectrometer \cite{gabalski2026fast}. The ions and electrons were accelerated by switchable electrostatic lenses onto a microchannel-plate (MCP) phosphor screen detector. The high voltages on the VMI electrodes were switched on the 10-ns timescale to allow coincident detection of electrons and ions from the same laser shot. This switching capability is essential for the CRATI measurement, since it enables the photoelectron spectrum and the ion fragmentation pattern to be recorded from identical molecules, preserving the channel-resolved information that would otherwise be lost in a single-particle detection scheme. The relative energy resolution of the electron spectrometer was $\Delta E/E \approx 2\%$ across the measured range, and the detector collection efficiency approached the full $4\pi$ solid angle. Calibration against the well-known above-threshold ionization (ATI) peaks of Argon was performed for the electron energy axis. The ion three-dimensional momenta were reconstructed from their time-of-flight and detector positions using the calibrated spectrometer fields.

Deuterated water (D$_{2}$O) was chosen because the heavier isotope slows the nuclear motion relative to H$_{2}$O, providing more time resolution for the electron--nuclear coupled dynamics. Due to natural hydrogen exchange, H$_{2}$O and HDO were also present in the sample and their ionization data were collected at the same time.

\newpage

%%%%%%%%%%%%%%%% SUPPLEMENTARY TEXT %%%%%%%%%%%%%%%
\subsection*{Supplementary Text}
\subsubsection*{Data analysis}

Photoelectron momentum images were recorded on the MCP-phosphor detector. The raw two-dimensional position data $(x,y)$ were converted to energy--angle-resolved spectra using the numerical Abel inversion algorithm. The energy calibration was verified against the ATI peak positions of Argon. The resulting three-dimensional momentum components $(p_{x}, p_{y}, p_{z})$ were obtained under the cylindrical symmetry assumption inherent to the VMI geometry. Ion three-dimensional momenta were reconstructed directly from the detector position $(x,y)$ and time-of-flight $t$ using the calibrated spectrometer fields. The relation $\mathbf{p} = F(x,y,t)$ maps each hit to a momentum vector in the lab frame, where the $x$-axis is aligned with the laser electric field and the $z$-axis coincides with the time-of-flight direction. For the pump--probe data, a momentum conservation filter $|\sum \mathbf{p}| \le 10$~a.u. was applied to retain only complete fragmentation events originating from the same parent molecule.

Covariance mapping \cite{frasinski2016covariance} was frequently used in understanding two- and three-body correlations. The extension to four- and more body correlation with cumulant mapping technique \cite{frasinski2022cumulant,cheng2023multiparticle} and recent development in computational algorithm \cite{cheng2024multiparticle} has greatly boosted the capability of the method. Electron--ion covariance maps were computed from single-shot coincidence data to isolate correlated photoelectron--photoion pairs from uncorrelated background. The joint probability $C(E_{e}, E_{\text{ion}})$ was obtained by correlating the electron energy spectrum with the kinetic energy of the detected ion on a shot-by-shot basis, effectively suppressing events where the electron and ion originated from different molecules. The 4th-order cumulant expansion was used to subtract uncorrelated background terms from the raw covariance, and the resulting cumulant maps were transformed into the molecular recoil frame to remove the laboratory-frame angular bias introduced by the laser polarization.

The kinetic energy release (KER) was calculated from the momentum magnitudes of the ion pairs: $\text{KER} = p_{1}^{2}/(2m_{1}) + p_{2}^{2}/(2m_{2})$. For the D$_{2}^{+}$--O$^{+}$ channel, the internuclear distance $R$ at the instant of the probe pulse was inferred from the KER using a two-point-charge Coulomb interaction approximation: $R = 1/\text{KER}$ (in atomic units). This mapping is accurate for two bare charges but becomes approximate at small separations ($R < 4$~a.u.) where the actual dicationic potential deviates from the pure Coulomb form due to electronic screening. The validity of this approximation is discussed in following section regarding the 8-fs enhancement feature.

Trajectory surface-hopping (TSH) simulations were performed in OpenMolcas at the CASSCF level of theory. The initial geometry was sampled from the $\tilde{B}$ state of D$_{2}$O$^{+}$ at a bent configuration ($\theta_{\text{DOD}} \approx 170^{\circ}$, $r_{\text{OD}} \approx 1.5$~\AA) with initial velocities drawn from a 300-K Boltzmann distribution. The nonadiabatic coupling vectors were computed on-the-fly, and surface hops were decided using the Tully fewest-switches algorithm. The time step was 20 atomic units ($\sim$0.5~fs). The 1$a_{1}$ and 2$a_{1}$ orbitals (predominantly O 1s and 2s) were kept doubly occupied throughout; the active space comprised the valence orbitals. The first three cationic potential-energy surfaces were calculated simultaneously for each trajectory.

\newpage

\subsubsection*{D$_{2}^{+}$ side-peak discussion}

In the CRATI photoelectron spectrum correlated with D$_{2}^{+}$ (Fig.~1(b) of the main text), weak additional peaks are observed near 4~eV and 7~eV that are absent from the parent D$_{2}$O$^{+}$ and D$^{+}$ spectra. These peaks are tentatively attributed to one of two mechanisms. (i) Direct ionization from the inner-valence 2$a_{1}$ orbital, which has an ionization potential of $\approx$~28.5~eV, followed by population of the $\tilde{C}$ state and subsequent dissociation into D$_{2}^{+}$. (ii) A resonant enhanced multiphoton ionization (REMPI) process involving higher-lying neutral or cationic states. The relative yield of this channel is small compared with the dominant A-state pathway. Unlike the main A-state pathway, these side peaks do not share the same photoelectron comb location as the parent D$_{2}$O$^{+}$ ion and are not correlated with the characteristic KER distribution of the (D$_{2}^{+}$, O$^{+}$) channel from the non-Franck-Condon dynamics. We therefore treat them as a secondary process that does not affect the main conclusions of this work. A definitive assignment of the 4-eV and 7-eV peaks would require photoelectron measurements with shorter-wavelength pulses (e.g., 200~nm) to resolve the higher-lying states, which is beyond the scope of the present study.

\newpage

\subsubsection*{8-fs enhancement: angular distribution and intensity threshold}

The pronounced enhancement at $\pm 8$~fs and KER~$\approx$~4.5~eV ($R \approx 2.5$~\AA) in the (D$_{2}^{+}$, O$^{+}$) channel belongs to the collective dissociation trajectory from the simulation. The following three sets of evidence support this assignment.

\begin{figure} % Do not use \begin{figure*}
	\centering
	\includegraphics[width=0.8\textwidth]{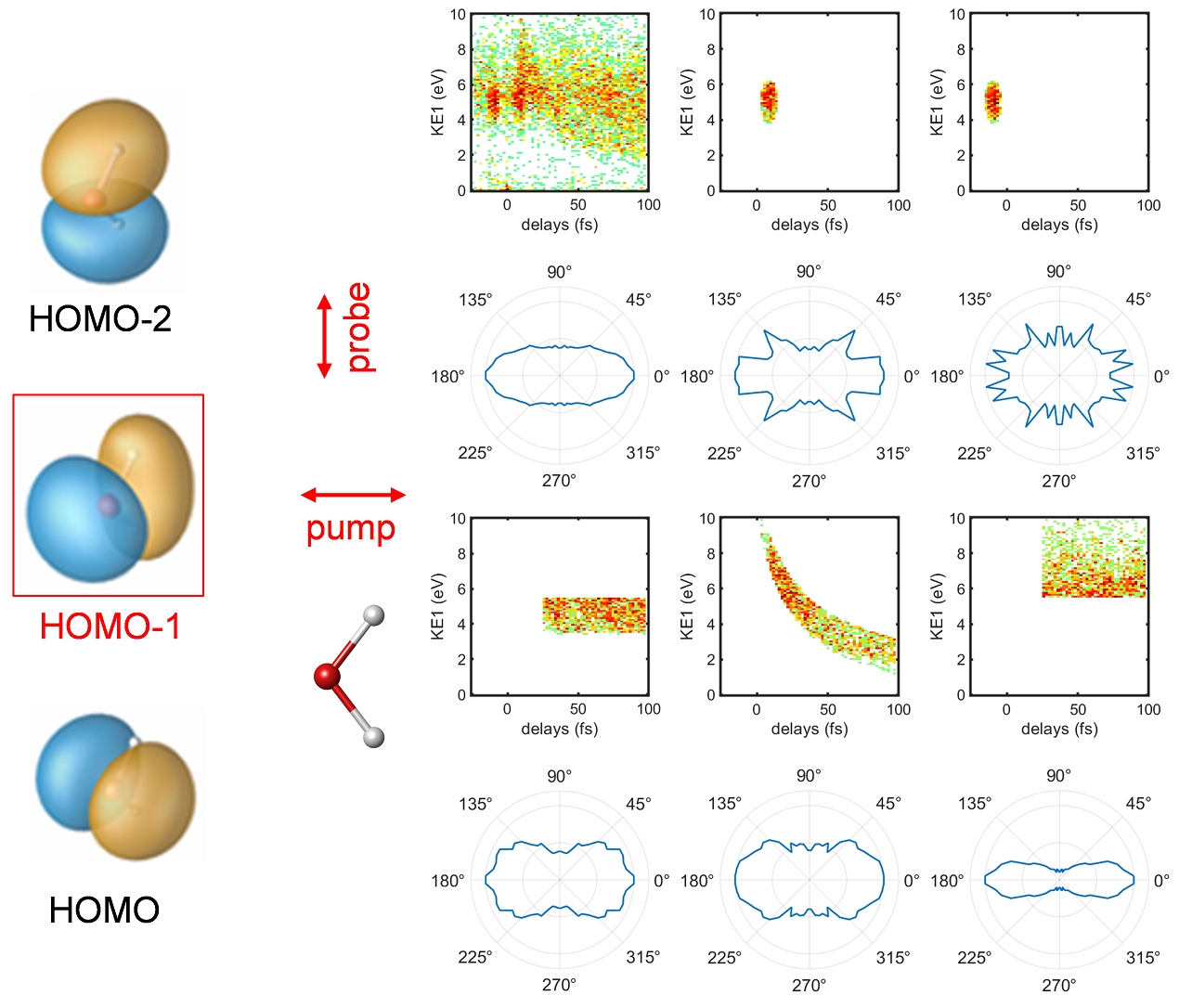}
	% Pick an appriopriate width for the size of the image

	% Captions go below figures
	\caption{\textbf{Angular distribution of (D$_{2}^{+}$, O$^{+}$) ion--ion correlations in the laboratory frame.}
		Left: Three valence electron orbital shapes. Middle: pump and probe pulse orientation in the lab frame. Right: Gated delay-KER plots and their corresponding angular distribution of (D$_{2}^{+}$, O$^{+}$). The gate on the enhancement region at $\pm 8$~fs and KER~$\approx$~4.5~eV shows signal preferentially aligns with the pump polarization along the O--D$_{2}$ axis ($0^{\circ}$ and $180^{\circ}$), supporting HOMO--1 ionization. Because the 3$a_{1}$ orbital has its maximum ionization amplitude along the molecular axis.}
	\label{fig:ang_dist} % give each figure a logical label name
\end{figure}

\textbf{Angular-distribution evidence.}
By gating the angular distribution of the D$_{2}^{+}$--O$^{+}$ coincidence events, we find that the enhancement region preferentially aligns with the pump polarization along the O--D$_{2}$ molecular axis (Fig.~\ref{fig:ang_dist}). In comparison, the dissociation band at higher KER (KER~$>$~6~eV) also exhibits angular preference, as expected for molecules that have already rotated and bent significantly before reaching the asymptotic dissociation geometry. This geometric preference is consistent with ionization from the HOMO--1 (3$a_{1}$) orbital, whose nodal plane is perpendicular to the molecular axis.

\begin{figure} % Do not use \begin{figure*}
	\centering
	\includegraphics[width=0.8\textwidth]{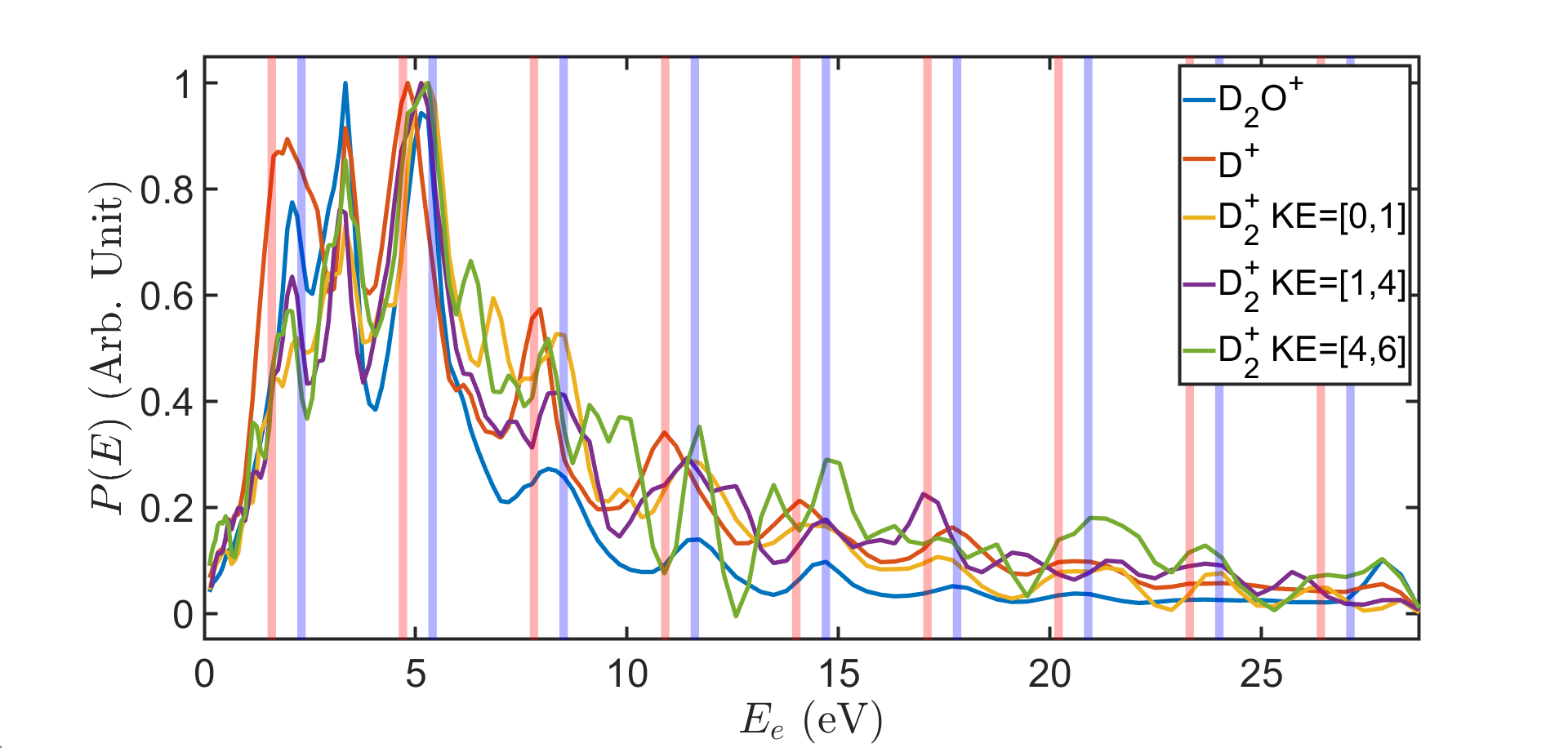}
	\includegraphics[width=0.8\textwidth]{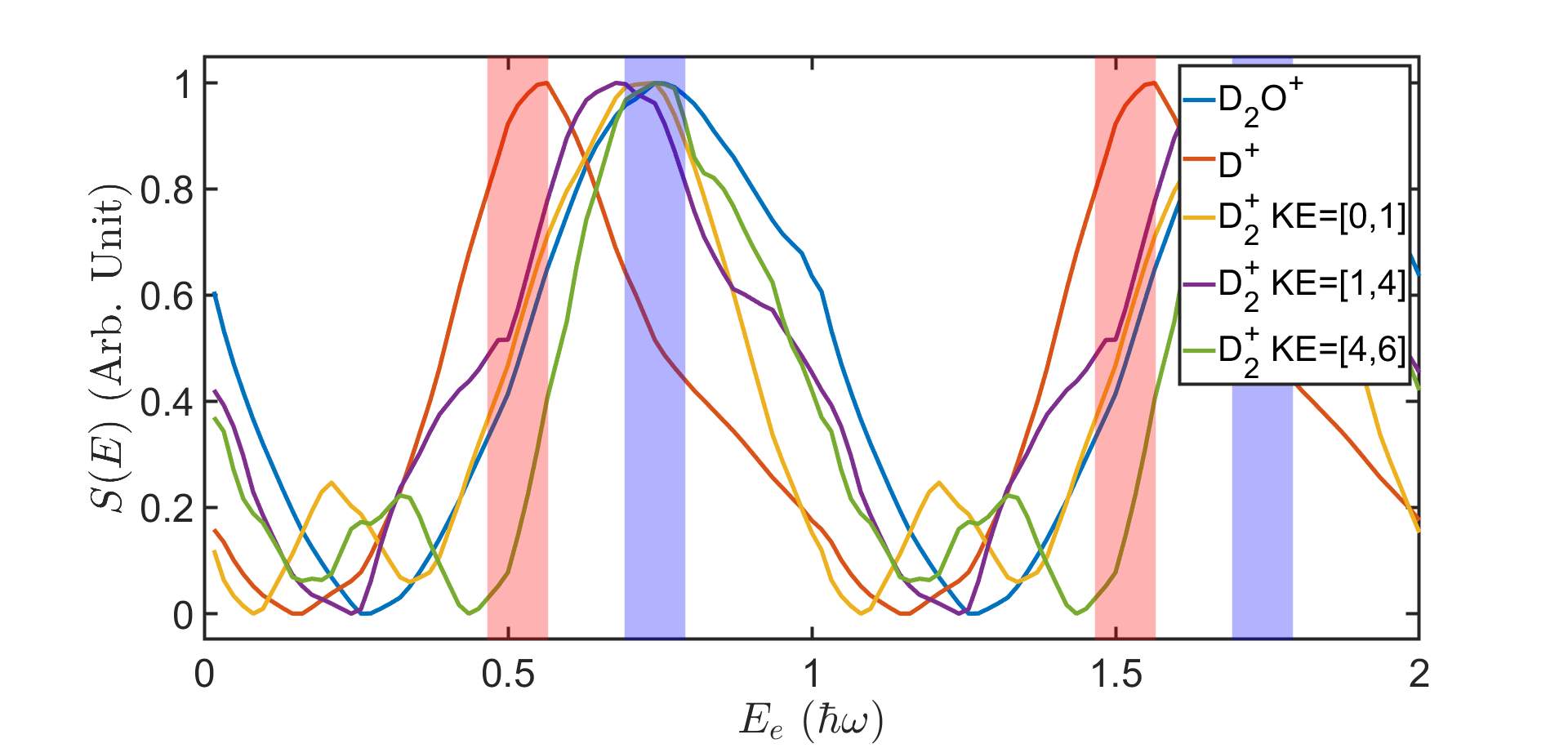}
	% Pick an appriopriate width for the size of the image

	% Captions go below figures
	\caption{\textbf{Momentum-resolved above-threshold ionization (MRATI) analysis of the 8-fs enhancement channel.}
		The momentum-resolved ATI analysis applied to KE-gated D$_{2}^{+}$ photoelectron spectra confirms that the 8-fs enhancement originates from the same A-state (3$a_{1}$)$^{-1}$ HOMO--1 ionization channel as the parent ion. The 4--6-eV comb region shows a clear HOMO--1 signature.}
	\label{fig:mrati} % give each figure a logical label name
\end{figure}

\textbf{MRATI evidence.}
Momentum-resolved ATI (MRATI) analysis was applied to the KE-gated D$_{2}^{+}$ photoelectron spectra to identify the ionization channel (Fig.~\ref{fig:mrati}). The KE-resolved spectra have limited statistics, so we used the comb-analysis method \cite{cheng2020momentum} to extract the ATI peak positions. All three KE-gated D$_{2}^{+}$ sets share the same dominant peak as the parent ion, i.e., the A-state (3$a_{1}$)$^{-1}$ HOMO--1 removal. The 4--6-eV comb region shows a clear HOMO--1 signature, confirming that the 8-fs enhancement channel originates from the same indirect ionization pathway as the bulk D$_{2}^{+}$ signal.

\begin{figure} % Do not use \begin{figure*}
	\centering
	\includegraphics[width=0.8\textwidth]{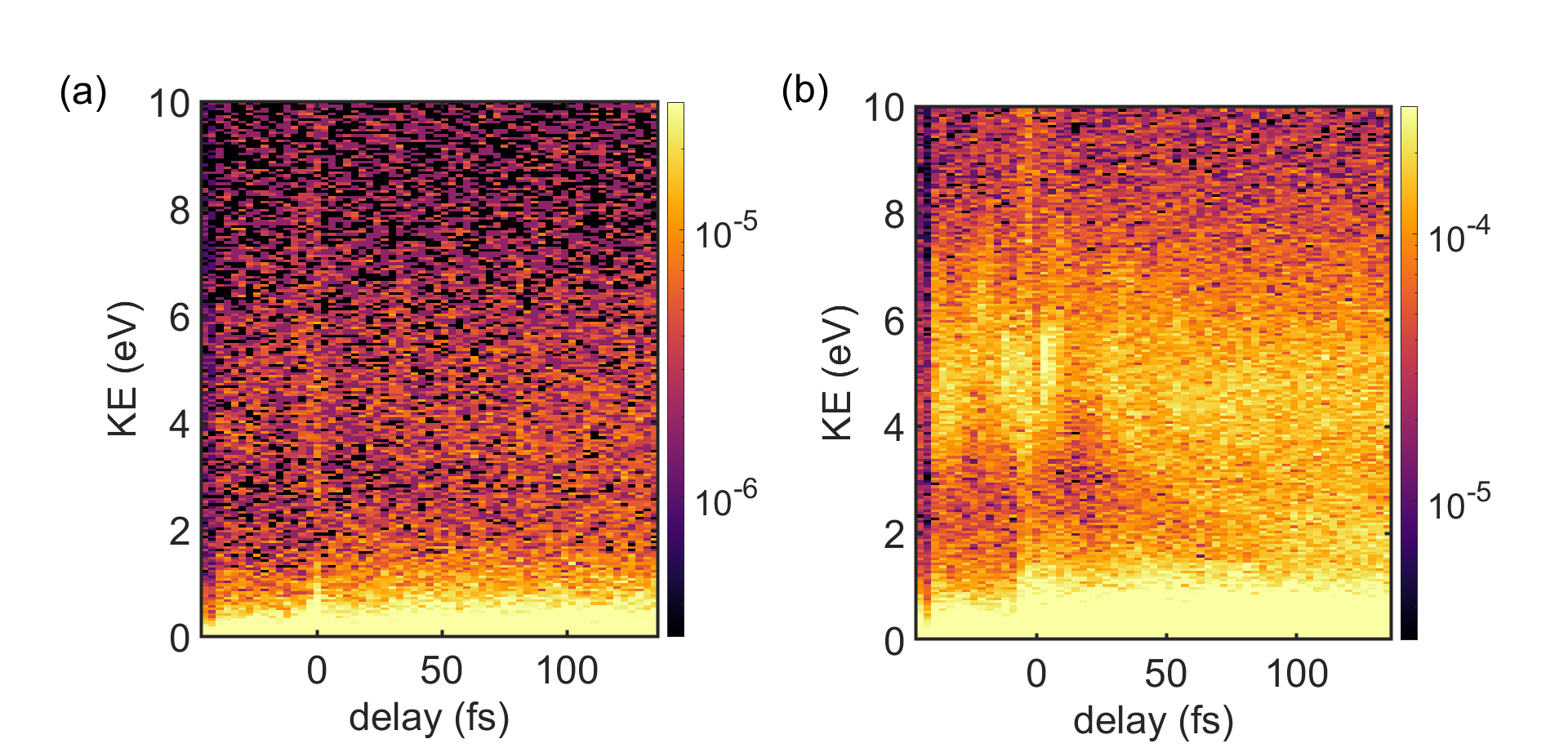}
	% Pick an appriopriate width for the size of the image

	% Captions go below figures
	\caption{\textbf{Intensity-dependent D$_{2}^{+}$ count rate as a function of kinetic energy and pump--probe delay.}
		Panel (a) was recorded at a probe intensity of $\approx$600~TW~cm$^{-2}$ and panel (b) at $\approx$1500~TW~cm$^{-2}$. The absence of the 8-fs enhancement at the lower intensity precludes a direct multiphoton single-ionization origin for the feature.}
	\label{fig:intensity} % give each figure a logical label name
\end{figure}

\textbf{Intensity-threshold evidence.}
An intensity-threshold argument further disfavors a direct double-ionization origin for the 8-fs enhancement. At probe intensities below $1\times10^{15}$~W~cm$^{-2}$, double ionization from each individual pulse is already observed in the D$_{2}^{+}$--O$^{+}$ channel, yet the 8-fs enhancement is completely absent (Fig.~\ref{fig:intensity}). If the enhancement were due to a direct multiphoton single-ionization followed up by double ionization process, it should scale with intensity and be present whenever double ionization occurs. Its absence at lower intensities indicates that the feature is not a simple delayed double ionization event. Focal-volume averaging complicates a clean intensity-dependent measurement, but the qualitative trend is clear.

\textbf{Interpretation of the 8-fs enhancement.}
Taken together, the angular distribution, MRATI, and intensity-threshold arguments are most consistent with a breakdown of the point-charge $R = 1/\text{KER}$ mapping at small internuclear separations. During the first $\sim$10~fs of non-Franck--Condon wave packet evolution, essentially all trajectories pass through compact geometries, where electronic screening makes the true dicationic potential much shallower than the pure Coulomb form. The KER then depends only weakly on the O--D$_2$ separation, so trajectories spanning a broad range of true internuclear distances acquire nearly the same KER and are compressed by the point-charge inversion into a narrow band of apparent $R$. The resulting pileup, rather than an enhanced dissociation probability at this geometry, produces the sharp feature in the reconstructed $R$--delay map. Contribution from double-ionization is also possible. The dicationic-state coupling discussed below provides one possible microscopic mechanism. Time-resolved photoelectron data would be required to distinguish between these contributions, but such experiments require short laser pulses for time resolution and short wavelength to spectrally resolve the features, which is beyond the scope of this work.

\textbf{Dicationic-state coupling.}
A genuine enhancement could arise from strong configuration mixing between dicationic states at the compact geometry probed at $\pm$8~fs ($r_{\text{OD}1} = r_{\text{OD}2} \approx 1.15$~\AA, DOD angle $\approx 135^{\circ}$), which is distinct from the nonadiabatic coupling at the conical intersection. In this geometry, the $^1$B$_2$ state of D$_2$O$^{2+}$, with nominal orbital occupancy $(1b_2)^1(3a_1)^1(1b_1)^2$, mixes strongly with the $^1$A$_1$ state. 
The resulting dipole-allowed transition between these mixed states provides an additional coupling during the probe pulse, enhancing the D$_2^+$ yield at this specific nuclear configuration. This state coupling is independent of the $\tilde{B}$--$\tilde{A}$ conical intersection and provides an alternative mechanism for population transfer that is resonant at the compact geometry probed at $\pm$8~fs. Because the dicationic states are highly sensitive to the O--D bond length, the enhancement is sharply localized in both delay and KER, producing the narrow feature observed in the $R$--delay map.

\newpage

\subsubsection*{Group of trajectories}

Out of 500 trajectory-surface-hopping trajectories launched from the $\tilde{B}$ state of D$_{2}$O$^{+}$, only 69 satisfy the D$_{2}^{+}$ selection criteria (D--D distance $<2$~\AA\ and O--D$_{2}$ center-of-mass distance $>1.8$~\AA\ at the end of the 100-fs propagation). This low fraction (13.8\%) reflects the stringent physical constraints on D$_{2}^{+}$ formation. The molecule must first undergo nonadiabatic population transfer from the $\tilde{B}$ to the $\tilde{A}$ state at the conical intersection, then execute the specific nuclear rearrangements that bring the two deuteriums into bonding proximity while simultaneously breaking both O--D bonds.

\begin{figure} % Do not use \begin{figure*}
	\centering
	\includegraphics[width=0.8\textwidth]{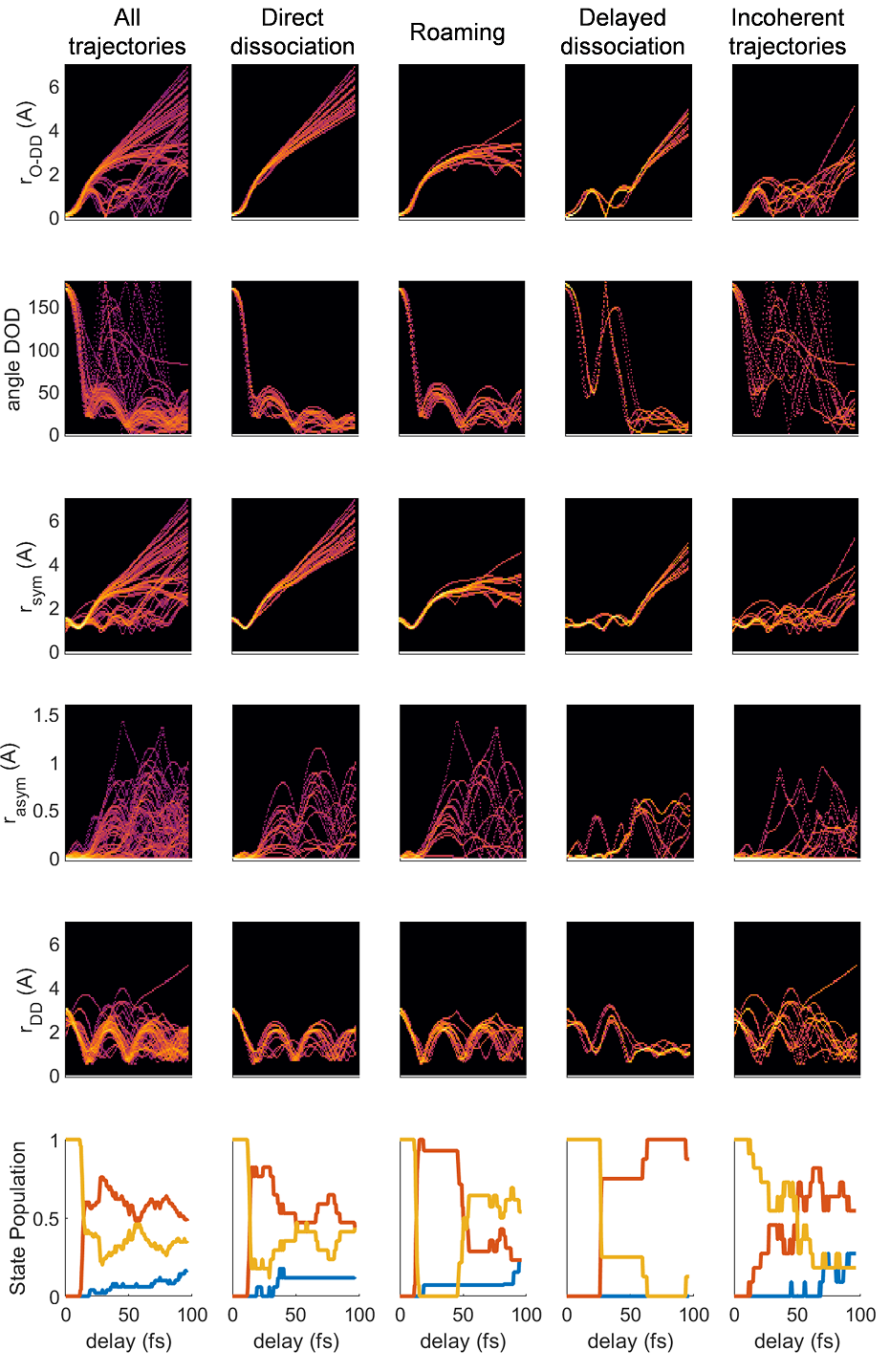}
	% Pick an appriopriate width for the size of the image

	% Captions go below figures
	\caption{\textbf{Classification of all 69 trajectories satisfying the D$_{2}^{+}$ selection criteria.}
		Direct dissociation (23 trajectories) hop at the first conical intersection encounter and dissociate monotonically; roaming (19 trajectories) execute large-amplitude O--D$_{2}$ oscillations before eventual dissociation; delayed dissociation (13 trajectories) remain on the upper adiabatic surface and dissociate after a second CI passage; unclassified (14 trajectories) exhibit irregular dynamics.}
	\label{fig:traj_groups} % give each figure a logical label name
\end{figure}

The 69 trajectories are classified into four dynamical groups based on their time-dependent behavior (Fig.~\ref{fig:traj_groups}).

(1)~\textit{Direct dissociation} (23 trajectories): these trajectories hop to the lower-lying $\tilde{A}$ surface at the first encounter with the $\tilde{B}$--$\tilde{A}$ conical intersection and proceed monotonically to D$_{2}^{+}$ + O. The O--D$_{2}$ distance increases steadily after the hop, with no significant turning points.

(2)~\textit{Roaming} (19 trajectories): these trajectories also hop at the CI but subsequently turn back toward the oxygen, executing large-amplitude D--D--O motion. The O--D$_{2}$ distance oscillates between $\sim$1.5 and $\sim$3.0~\AA, indicating that one deuterium temporarily orbits the other before eventual dissociation.

(3)~\textit{Delayed dissociation} (13 trajectories): these trajectories remain on the upper $\tilde{B}$ adiabatic surface after the first encounter with the CI, bounce off a high dissociation barrier, and dissociate adiabatically after a second passage through the CI region. The delay between the first and second CI encounters is typically 20--40~fs.

(4)~\textit{Unclassified} (14 trajectories): these trajectories exhibit irregular dynamics with multiple turning points and no clear classification into the above categories. Some remain adiabatic on the upper $\tilde{B}$ surface for an extended period before a late hop and subsequent dissociation; others show large-angle rotational motion that does not follow the typical bend--unbend pattern of the delayed trajectories, with the DOD angle increasing rather than decreasing after the initial CI encounter; still others roam without a clear pattern before eventually dissociating.

Because the initial conditions were sampled from a single fixed geometry (DOD angle $170^{\circ}$, $r_{\text{OD}} = 1.5$~\AA) with only thermal velocities varied, the observed spread in dynamical outcomes is primarily due to the stochastic nature of nonadiabatic surface hopping rather than a broad distribution of starting geometries. This means that the 69/500 fraction should be interpreted as a lower bound on the true D$_{2}^{+}$ branching ratio, and the quantitative discrepancy with the experimental yield is expected rather than problematic. The qualitative agreement (the existence of three distinct dissociation modes and the correct order of magnitude of the time constants) is the relevant validation criterion.

Fig.~\ref{fig:traj_groups} displays all 69 trajectories classified by group. The direct-dissociation trajectories form the fastest-rising population, the roaming trajectories contribute the delayed component, and the delayed-dissociation trajectories bridge the two timescales. The unclassified trajectories are scattered throughout the map and do not contribute a coherent signal.

\newpage

\subsubsection*{Fitting time constants for dissociation}

\begin{figure} % Do not use \begin{figure*}
	\centering
	\includegraphics[width=0.8\textwidth]{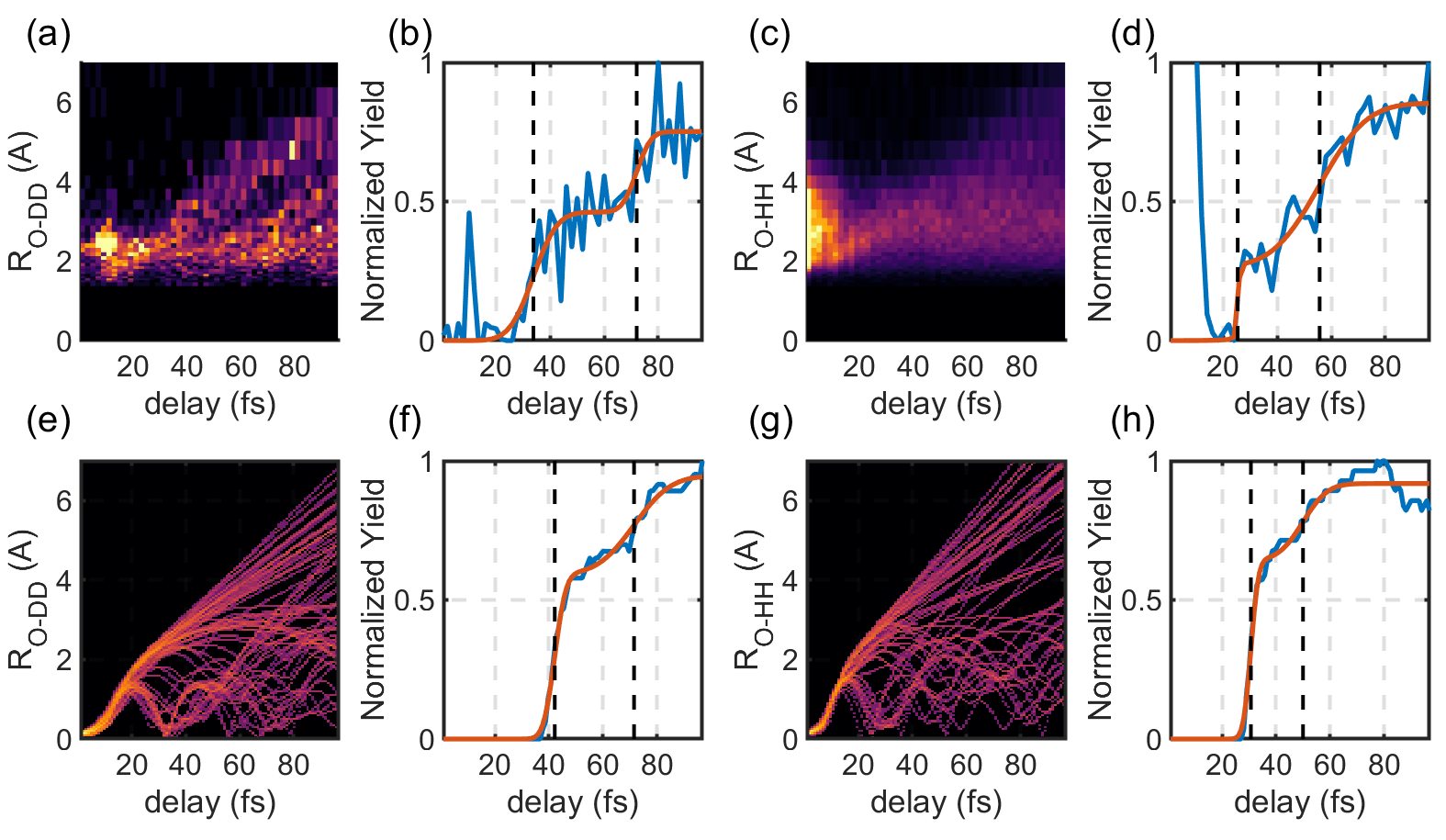}
	% Pick an appriopriate width for the size of the image

	% Captions go below figures
	\caption{\textbf{Error-function fits to the experimental and simulated dissociation time constants.}
		(a-d)~Experiment. Reconstructed O--D$_2$ and O--H$_2$ distance--delay maps for the (D$_{2}^{+}$, O$^{+}$) (a) and (H$_{2}^{+}$, O$^{+}$) (c) coincidence channels, and the corresponding delay-dependent yields (blue) fitted with two sequential error functions (orange) (b,d). Dashed lines mark the fitted rise centers at $\mu_1 = 33.8$~fs and $\mu_2 = 72.0$~fs for D$_{2}^{+}$, and $\mu_1 = 25.4$~fs and $\mu_2 = 55.7$~fs for H$_{2}^{+}$. The isotope ratio $\mu_1$(D$_{2}^{+}$)/$\mu_1$(H$_{2}^{+}$) $\approx$ 1.33 is consistent with the square-root mass scaling expected for nuclear motion on the same potential-energy surface.
		(e-h)~Same analysis for the simulated trajectory ensembles, giving $\mu_1 = 42.2$~fs and $\mu_2 = 71.5$~fs for D$_{2}^{+}$ and $\mu_1 = 30.7$~fs and $\mu_2 = 50.0$~fs for H$_{2}^{+}$. All fit parameters are listed in Table~\ref{tab:fit_params}.}
	\label{fig:fit_time} % give each figure a logical label name
\end{figure}

To extract the time constants for D$_{2}^{+}$ formation, the time-dependent yield was constructed from both the experimental and simulated data. Fig.~\ref{fig:fit_time} shows the fitted curves for both D$_{2}^{+}$ and H$_{2}^{+}$. The top panel of Fig.~\ref{fig:fit_time} shows the two independent error-function fits for D$_{2}^{+}$, while the bottom panel shows the corresponding fits for H$_{2}^{+}$. And the fitted populations and time constants are summarized in Table~\ref{tab:fit_params}. For the experiment, the $r$--delay map was constructed from the KER--delay data using $r = 1/\text{KER}$ (in atomic units). The yield was then integrated over the dissociation region ($r > 2.6$~\AA) to obtain a one-dimensional curve. For the simulation, the same procedure was applied to the $R$--delay map from the TSH trajectories, integrating over $R > 2.6$~\AA. Both curves were fitted with a sum of two complementary error functions (Eq.~S1),
\begin{equation}
Y(t) = A_{1}\left[1 + \text{erf}\left(\frac{t - \mu_{1}}{\sigma_{1}}\right)\right] + A_{2}\left[1 + \text{erf}\left(\frac{t - \mu_{2}}{\sigma_{2}}\right)\right],
\label{eq:fit}
\end{equation}
representing the two sequential rise components (direct and delayed dissociation). For the experimental data, the first rise is centered at $\mu_{1} = 33.8$~fs with width $\sigma_{1} = 9.8$~fs, and the second rise at $\mu_{2} = 72$~fs with $\sigma_{2} = 5.9$~fs. For the simulation, the corresponding parameters are $\mu_{1} = 42.2$~fs, $\sigma_{1} = 4.6$~fs and $\mu_{2} = 71.5$~fs, $\sigma_{2} = 15.2$~fs. The experimental second rise is narrower ($\sigma_{2} = 5.9$~fs) than the simulated one ($\sigma_{2} = 15.2$~fs), indicating that the delayed dissociation pathway is more uniform in the experiment than in the simulation. The broader simulated distribution reflects a wider range of barrier heights and reflection times in the simplified theoretical model, whereas the experiment likely averages over a more restricted subset of delayed trajectories due to the finite probe pulse duration and the focal volume. 

For comparison, the same procedure was applied to the (H$_{2}^{+}$, O$^{+}$) channel. There are 200 trajectories for the H$_{2}$O simulation and 48 of them ended up with (H$_{2}^{+}$, O$^{+}$) dissociation channel. In the experiment, the first rise occurs at 25.4~fs with $\sigma = 1.0$~fs, consistent with the lighter isotope's faster nuclear motion. The isotope ratio of the first rise times ($33.8/25.4 \approx 1.33$) is close to the square root of the mass ratio ($\sqrt{2} \approx 1.41$), as expected for hydrogen-bond dissociation dynamics. The H$_{2}^{+}$ channel shows the same two-rise structure as D$_{2}^{+}$, but the second rise is broader ($\sigma_{2} = 19.4$~fs) and less well-defined. This broader width reflects the faster nuclear motion of H$_{2}^{+}$, which causes the delayed-dissociation trajectories to sample a wider range of barrier heights and reflection times before exiting to the dissociation limit. The second rise in H$_{2}^{+}$ is therefore less uniform than in D$_{2}^{+}$, consistent with the mass-dependent sensitivity of the delayed pathway to the initial conditions and the probe pulse duration. The faster timescale of H$_{2}^{+}$ also means that the pump--probe delay window ($-20$~fs to $+100$~fs) captures a larger fraction of the total dissociation dynamics, reducing the relative weight of the unobserved long-time tail and making the second rise appear earlier and less distinct.

%%%%%%%%%%%%%%%% SUPPLEMENTARY FIGURES %%%%%%%%%%%%%%%

%%%%%%%%%%%%%%%% SUPPLEMENTARY TABLES %%%%%%%%%%%%%%%

\begin{table} % Do not use \begin{table*}
	\centering
	% Captions go above tables
	\caption{\textbf{Fitted time constants for D$_{2}^{+}$ and H$_{2}^{+}$ formation.}
		Summary of error-function fit parameters for the two sequential rise components extracted from the experimental data.}
	\label{tab:fit_params} % give each figure a logical label name

	\begin{tabular}{lcccc} % four columns, alignment for each
		\\
		\hline
		Channel & $\mu_1$ (fs) & $\sigma_1$ (fs) & $\mu_2$ (fs) & $\sigma_2$ (fs)\\
		\hline
		D$_{2}^{+}$(exp) & 33.8 & 9.8 & 72.0 & 5.9\\
		D$_{2}^{+}$(sim) & 42.2 & 4.6 & 71.5 & 15.2\\
		H$_{2}^{+}$(exp) & 25.4 & 1.0 & 55.7 & 19.4\\
		H$_{2}^{+}$(sim) & 30.7 & 3.1 & 50.0 & 10.2\\
		\hline
	\end{tabular}
\end{table}

%%%%%%%%%%%%%%%% END OF SUPPLEMENT %%%%%%%%%%%%%%%

\end{document}